\documentclass[5p]{elsarticle}

\usepackage{lineno,hyperref}
\usepackage{amsmath}
\usepackage{amssymb}
\usepackage{xspace}
\modulolinenumbers[5]
\usepackage{graphicx}
\usepackage{bm}
\usepackage[justification=raggedright]{caption}
\usepackage{subcaption}
\usepackage{physics}
\usepackage{algorithm}
\usepackage{algorithmic}
\usepackage{comment}
\usepackage{xcolor} 
\usepackage{hyperref}
\usepackage{empheq}

\newcommand{\pp}[1]{{#1}_{\parallel}}

\newcommand{\rcy}{\mathfrak{R}}
\newcommand{\rev}[1]{{\color{black}#1}}
\newcommand{\revv}[1]{{\color{black}#1}}

\journal{Nuclear Materials and Energy}

\begin{document}

\begin{frontmatter}
\title{Reduced Physics Model of the Tokamak Scrape-off-Layer for Pulse Design}
\author[PPPL]{X. Zhang\corref{mycorrespondingauthor}}
\ead{xzhang@pppl.gov}
\cortext[mycorrespondingauthor]{Corresponding author}
\author[PPPL]{F. M. Poli}
\author[PPPL]{E. D. Emdee}
\author[PPPL]{M. Podest\`{a}}
\author{the NSTX-U Team}

\address[PPPL]{Princeton Plasma Physics Laboratory, 100 Stellerator Rd, Princeton, NJ 08536, USA}

\begin{abstract}
The dynamic interplay between the core and the edge plasma has important consequences in the confinement and heating of fusion plasma. The transport of the Scrape-Off-Layer (SOL) plasma imposes boundary conditions on the core plasma, and neutral transport through the SOL influences the core plasma sourcing. In order to better study these effects in a self-consistent, time-dependent fashion with reasonable turn-around time, a reduced model is needed. In this paper we introduce the SOL Box Model, a reduced SOL model that calculates the plasma temperature and density in the SOL given the core-to-edge particle and power fluxes and recycling coefficients. The analytic nature of the Box Model allows one to readily incorporate SOL physics in time-dependent transport solvers for pulse design applications in the control room. Here we demonstrate such a coupling with the core transport solver TRANSP and compare the results with density and temperature measurements, obtained through Thomson scattering and Langmuir probes, of an NSTX discharge. Implications for future interpretive and predictive simulations are discussed.
\end{abstract}


\end{frontmatter}

\section{Introduction}
The dynamic interplay between the core and the edge plasma has important consequences in the confinement and heating of fusion plasmas. \revv{The effects of the Scrape-Off-Layer (SOL) on the core confined plasma can be significant in many contexts, including, but not limited to, neutral beam deposition \cite{garcia2013fast, kramer2013simulation}, radio-frequency heating and current drive \cite{bertelli2015full, perkins2013fast, poli2021implications}, impurity transport \cite{gao2020plasma, morozov2007impurity}, plasma fueling \cite{mordijck2020overview}, and neutral friction and electric field formation \cite{rozhansky2001simulation}.} To systematically study these problems and enable control-room applications, we need to be able to simulate the entire tokamak plasma self-consistently and time-dependently with reasonable turn-around time. Reduced physics models are well-suited for this purpose because of their lighter computational cost and the ability to reasonably reproduce physical effects within their validated parameter spaces.

The Two-Point Model of SOL plasma is a widely used 0-D stationary model of SOL transport. It consists simply of conservation equations that relate the quantities found at ``upstream", typically defined as the outer mid-plane (OMP) of the tokamak or sometimes the divertor entrance, to those found at the divertor targets. However, the model requires the upstream density as an input parameter so that the other plasma parameters can be calculated, thus limiting its ability as a predictive model.
Various efforts have been made in calculating this upstream density via other known quantities to achieve a predictive Two-Point Model, adopting either a database approach \cite{stangeby2021role, lore2021model} or a diffusive ansatz in the cross-field direction with chosen scale-lengths \cite{siccinio20160d}. The former approach is inherently device-specific, while the latter approach can over-estimate the upstream density since the slow cross-field transport is the only particle sink. \revv{Ref.~ \cite{luda2020integrated} adopts an analytical formula derived in \cite{kallenbach2019neutral} to calculate upstream density based on the divertor neutral pressure, which is needed as input to the model. In predictive simulations, this pressure, as well as the parameter that describes the effects of momentum exchange, power losses, and flux expansion, are found through regression analysis on the AUG database.}

We will show in this paper that by introducing a new picture for particle balance we can avoid the problem of an undetermined upstream density and instead calculate all SOL parameters from core power and particle fluxes only. Specifically, by balancing the cross-field transport of particles from core to SOL with the pumping at the targets, the plasma density can be calculated at both upstream and the target. Recycling coefficients are required as additional control parameters (Sec.~\ref{sec:modBox}). When coupled to a transport solver like TRANSP \cite{hawryluk1981empirical, doecode_12542}, the fluxes are calculated by the core transport solver and the recycling coefficients can be constrained using divertor measurements, as will be demonstrated in Sec.~\ref{sec:NSTX}. This approach will offer new opportunities for efficient core-edge coupled predictive modeling as well as interpretive analysis of target recycling processes.

\section{Derivation of the 0-D Box Model}
\label{sec:modBox}

\begin{figure}
    \centering
    \includegraphics[width = \linewidth]{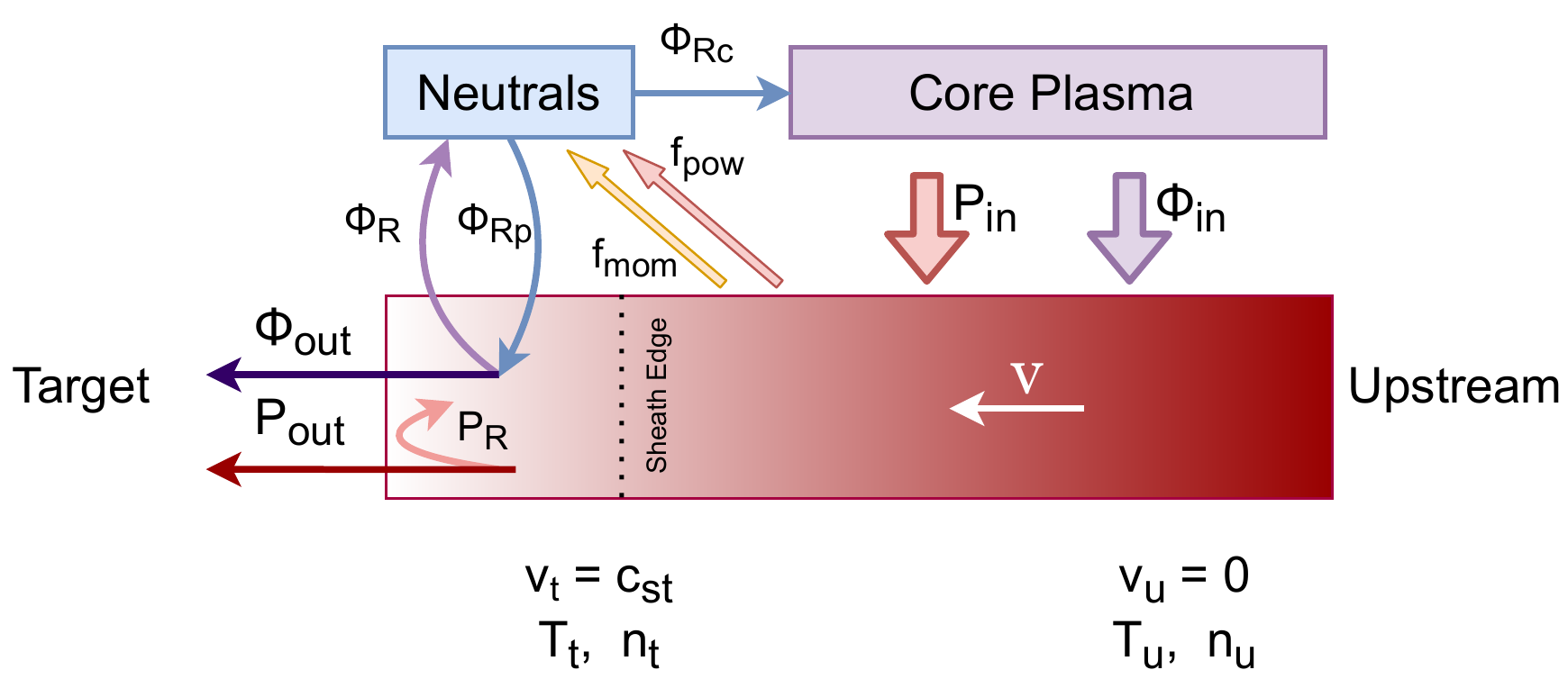}
    \caption{Schematic of the box model particle and energy balance, where $P$ is power, $\Phi$ is particle flux, and gray line marks the sheath edge. Particle and energy flow between the 3 boxes: the core plasma, the SOL plasma, and neutrals. To achieve steady state in the SOL, we must have $P_{in} = P_{out}$ and $\Phi_{in} = \Phi_{out}$.}
    \label{fig:extended}
\end{figure}

The 0-D Box Model is an extension of the conduction-limited Two-Point Model \cite{stangeby2000plasma}, which explicitly accounts for the balance between core to SOL particle flux and target pumping. A schematic of the particle and power balance picture is shown in  Fig.~\ref{fig:extended}. Particles and energy are assumed to flow among three ``boxes": the core plasma, the SOL plasma, and the neutrals. All input power $P_{in}$ and particle fluxes $\Phi_{in}$ (number of particles per second) into the core plasma are assumed to enter the SOL and are then transported via fluid flow towards the target. At the target surface, a fraction of the power $P_R$ is returned to the SOL plasma while a fraction $P_{out}$ gets transmitted out of the system via the plasma sheath. 

Ions that hit the target are either removed from the system at rate $\Phi_{out}$ through surface retention or active pumping, or are recycled as neutrals at rate $\Phi_R$. These neutral particles re-enter the SOL either through direct ionization in the SOL at rate $\Phi_{Rp}$, or via core-to-SOL transport (governed by $\Phi_{in}$) after having been ionized in the core plasma at rate $\Phi_{Rc}$. Electrons that reach the material surface are either absorbed by the material at rate $\Phi_{out}$, or are returned to the plasma via reflection or secondary emission, or are simply re-created when recycled neutrals are ionized.

Similar to the conduction-limited Two-Point Model \cite{stangeby2000plasma}, the SOL plasma is assumed to be a fluid, with Braginskii-like thermal conductivity. Plasma pressure is also assumed to be conserved along the flux tube. This leads to the first two equations of the Box Model, which are identical to those in the standard Two-Point Model \cite{stangeby2000plasma}:
\begin{align}
    n_u (kT_u^i + kT_u^e) &= 2n_t(kT_t^i + kT_t^e),\label{eq:a}\\
    \qty(T_u^{s})^{7/2} &= \qty(T_t^{s})^{7/2} + \frac7{4} \frac{\pp{q}^s L}{\kappa_{0s}},\label{eq:b}
\end{align}
where the subscript $u$ indicates ``upstream", which is assumed to be at the OMP, the subscript $t$ denotes ``target", the superscript $s$ denotes particle species, $n$~[m$^{-3}$] is plasma density, $T$~[eV] is plasma temperature, $L$~[m] is the SOL parallel connection length, $\pp{q}^s= P_{in}^s/\pp{A}$ is the parallel heat flux, $P_{in}$ the total power into the SOL, $\pp{A}$ the parallel area of the SOL, and $\kappa_{0s}$ is the coefficient of electron and ion conductivity as defined in \cite{stangeby2000plasma}. 


In contrast to the complete particle recycling of the conduction-limited Two-Point Model, in the Box Model the plasma sheath in front of the target is both an energy and particle sink. Consistent with the spatially uniform power input to the SOL, we start by assuming that the particle fluxes into the SOL are also spatially uniform:
\begin{equation}
    S_{pc}^s = \frac{\Gamma^s}{L} \equiv  \frac{\Phi_{in}^s}{A_{||} L}
\end{equation}
where $S_{pc}$~[m$^{-3}$s$^{-1}$] is the particle source from the core, $\Gamma^s$~[m$^{-2}$s$^{-1}$] is the particle flux density, and $\Phi_{in}^s$~[s$^{-1}$] is the particle flux from the core into the SOL.

In order for the SOL to maintain steady state, we must now introduce ``leakage" via target pumping to balance the particle input. We define a particle transmission rate as $\gamma_{part}^s = 1 - \rcy^s$, where $\rcy^s$ is the effective particle recycling coefficient. For simplicity, all recycled neutrals are from now on assumed to be ionized in the SOL, close to the target. This assumption is better suited in diverted geometries where the target is far away from the plasma core, such as in the case of closed divertors, and when cross field transport in the SOL is weak such that only a negligible fraction of ions are lost to the main chamber compared to the divertors. The total effective electron recycling rate  $\rcy^e$ includes reflection, secondary emissions, and electrons created through ionization of recycled neutrals. 

Assuming that the plasma is accelerated to the sound speed in the sheath according to the Bohm criterion, global particle balance therefore implies that:
\begin{align}\label{eq:c}
    \gamma_{part}^s n_t c_{st} = \Gamma^s,
\end{align}
where $c_{st}$~[m~s$^{-1}$] is the target sound speed:
\begin{equation}
        c_{st} = \sqrt{\frac{kT_t^i + kT_t^e}{m_i}}\label{eq:cst}.
\end{equation}

The power balance equation now needs to be considered. To account for the recycled electrons, the sheath cooling power is modified by including the acceleration of the recycled electrons by the sheath potential:
\begin{align}
    \frac{P_{out}}{\pp{A}} &= \gamma_0 kT^e_t n_t c_{st}- (1 - \gamma_{part}^e)\abs{e V_p}n_t c_{st}\nonumber \\
    &= (2.5 + \alpha\gamma_{part}^e)kT^e_t n_t c_{st}
\end{align}
where $V_p$ is the plasma potential and $\gamma_0$ is the electron sheath heat transmission factor defined in \cite{stangeby2000plasma}:
\begin{align}
    \gamma_0 = 2.5 + \alpha,\quad
    \alpha  \equiv \frac{e V_p}{\abs{kT_t^e}}.
\end{align}

For simplicity, all ions are assumed to be recycled as cold neutrals with negligible energy. Since the plasma sheath has no effect on these recycled cold neutrals, the ion sheath heat transmission factor remains unchanged from the value assumed in the standard Two-Point Model, i.e. $\gamma_q^i \sim 2$. However, when the fraction of hot recycled neutrals is high, this sheath transmission factor would need to be modified accordingly to account for the energy returned to the SOL by the warm and hot neutrals. By power balance $P_{in} = P_{out}$ and assuming, for now, that the area of the flux tube stays constant, we arrive at the new power balance equation:
\begin{equation}\label{eq:d}
    \pp{q}^s = \gamma_q^s n_t c_{st} kT_{t}^s,
\end{equation}
where the revised electron sheath heat transmission factor is defined as $\gamma_q^e \equiv 2.5 + \alpha\gamma_{part}^e$.

Equations \ref{eq:a}, \ref{eq:b}, \ref{eq:c}, and \ref{eq:d} now form the complete set of equations for the 0-D Box model.
This system can now be exactly solved given $P^s_{in}$ and $\Phi_{in}^s$ as the only inputs, and $\pp{A}$, $L$ and recycling coefficients as control parameters.

\begin{figure}
    \includegraphics[width = 0.8\linewidth]{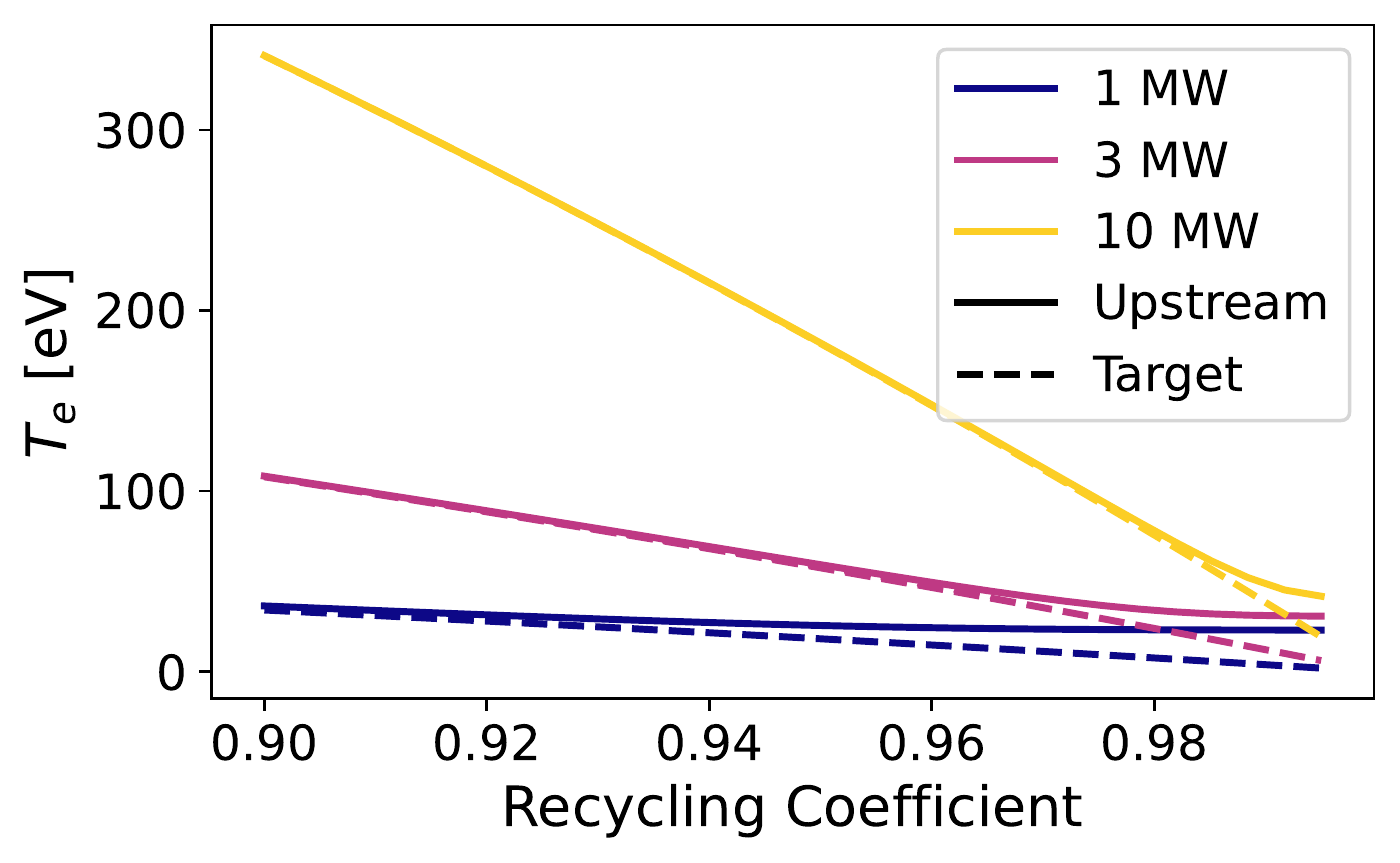}\\
    \includegraphics[width = 0.8\linewidth]{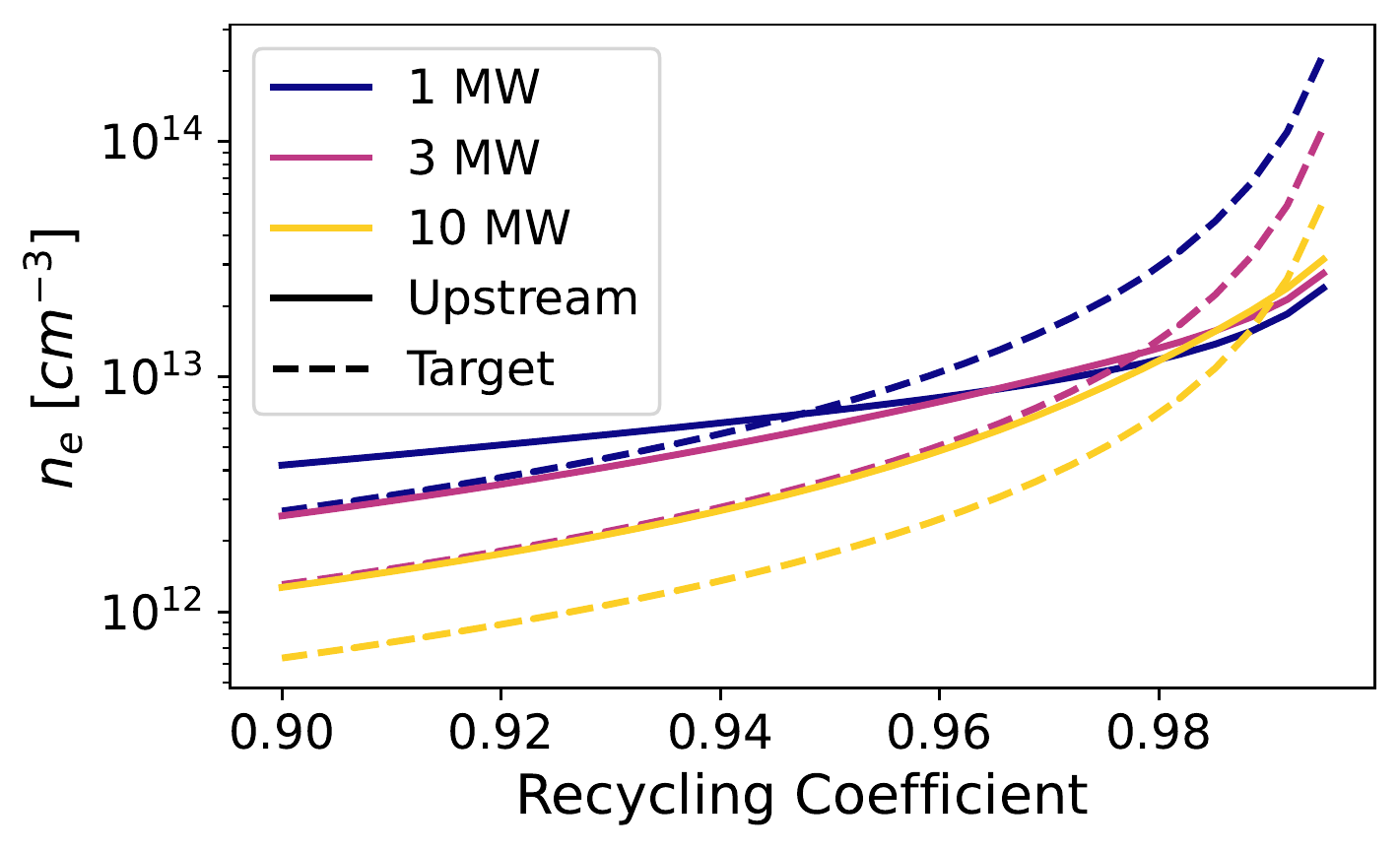}
    \caption{Calculated electron temperature and density as a function of recycling coefficient, assuming $\gamma_p^e = \gamma_p^i$. Connection length is set to 16~m and scrape-off-width to 9~mm, consistent with NSTX parameters. Core-to-SOL particle flux is fixed at $4\times10^{21}s^{-1}$. Three different input powers are shown. The SOL becomes isothermal as recycling decreases. Target density gradually becomes lower than upstream as recycling decreases. Both behaviors are consistent with the transition from conduction-limited to sheath-limited regimes.}
    \label{fig:TNofRho}
\end{figure}

Figures \ref{fig:TNofRho} show the dependency of calculated electron density and temperature on the specified recycling coefficients. NSTX-like parameters are used to calculate the connection length, and SOL power scrape-off-width is calculated according to \cite{goldston2011heuristic}. Consistent with expectations, as recycling is decreased, the SOL becomes increasingly isothermal while the target density becomes lower than upstream, \rev{signaling a smooth transition from the sheath limited regime at low recycling to the conduction limited regime at high recycling. The plasma density is overall lowered due to lower recycling, and plasma temperature is in turn raised from the decrease in dilution cooling.}


\subsection{Modifications to the 0-D Box Model}

The Two-Point Model assumptions of no volumetric loss of power and momentum and no flux expansion have been improved upon in recent years, forming the so-called ``Modified Two-Point Model" \cite{petrie2013effect, moulton2017using, kotov2009two}. These modification factors can similarly be included in the Box Model. Loss factors for volumetric power and momentum loss, $f_{pow}$ and $f_{mom}$ (due to collisions with neutrals, atomic processes, and viscous forces), along with the conduction fraction $f_{cond}$ have been derived in \cite{stangeby2000plasma} and will not be re-derived here. Modifications to the particle balance equation for finite flux expansion $f_R$ is similar to that in \cite{petrie2013effect}. The final resulting equations are as follows:
\begin{align}
    n_u (kT_u^i + kT_u^e) &= \frac{2}{f_{mom}}n_t(kT_t^i + kT_t^e),\label{eq:ma}\\
    \qty(T_u^{s})^{7/2} &= \qty(T_t^{s})^{7/2} 
          + f_{cond}\frac7{4} \frac{\pp{q}^s}{\kappa_{0s}}\frac{L\log(f_R)}{f_R - 1},\label{eq:mb}\\
    \pp{q}^s &= \frac{f_R}{1 - f_{pow}}\gamma_q^s n_t c_{st} kT_{t}^s,\label{eq:mc}\\
    \Gamma^s &= \frac{2 f_R}{f_R + 1}\gamma^s_{part} n_t c_{st},\label{eq:md}\\
    c_{st} &= \sqrt{\frac{kT_t^i + kT_t^e}{m_i}}.\label{eq:me}
\end{align}

The case of no volumetric losses corresponds to $f_{mom} = 1$, $f_{cond} = 1$, and $f_{pow} = 0$. Note that the loss factors do not influence the particle balance because the particle sources and sinks remain the same. We can now solve for all temperatures and densities in terms of known quantities:
\begin{align}
    kT_t^s &= \frac{2\gamma_{part}^s}{\gamma_q^s}\frac{1 - f_{pow}}{f_R + 1}\frac{\pp{q}^s}{\Gamma^s}\label{eq:modF_a}\\
    \qty(T_u^{s})^{7/2} &= \qty(T_t^{s})^{7/2} 
          + f_{cond}\frac7{4} \frac{\pp{q}^s}{\kappa_{0s}}\frac{L\log(f_R)}{f_R - 1}\label{eq:modF_b}\\
    n_t &= \frac{(f_R + 1)}{2f_R}\frac{\Gamma^s}{\gamma_{part}^s c_{st} }\label{eq:modF_c}\\
    n_u &= \frac{f_R + 1}{f_R f_{mom}}\frac{T_t^i + T_t^e}{T_u^i + T_u^e}\frac{\Gamma^s}{\gamma_{part}^s c_{st}}.\label{eq:modF_d}
\end{align}

Notably, compared to the results from \cite{petrie2013effect} where the upstream density $n_u$ is taken as the control parameter instead of the particle flux, the effect of flux expansion on target quantities is much weaker in the Box Model. Given recycling and input particle flux, the Box Model predicts
\begin{align}
    n_t \propto \frac{f_R + 1}{2f_R}, \quad
    T_t \propto \frac{2}{f_R + 1}.
\end{align}
The influence of flux expansion on both target quantities is therefore only significant when $f_R < 1$, i.e. when the flux is pinched near the target instead of expanded. The effect of $f_R$ diminishes quickly at values greater than 1\revv{, especially for the target density}. In contrast with the $n_t \propto f_R^2$ and $T_t \propto f_R^{-2}$ dependency reported in \cite{petrie2013effect}, this suggests that expanding flux is only marginally effective in lowering target temperature, with a factor of 2 increase in $f_R$ only leading to a factor of 0.67 reduction in temperature. The dependency of target density on $f_R$ is in fact reversed, with higher $f_R$ leading to lower target density, although the effect may be marginal at high $f_R$. These differences in dependency are a direct result of maintaining the particle flux as a constant, as opposed to the upstream density, \revv{and may help explain the weaker than expected dependency of detachment access on flux expansion reported in \cite{theiler2017results} - although further detailed transport studies are needed to confirm this hypothesis}. This suggests that flux expansion and compression have complex dependencies on the details of plasma fueling and density control, and warrant further studies with high fidelity simulations and dedicated experiments.

\rev{
\subsection{Limitations and Applicability of the Box Model}
Before moving on to an example application of the 0D Box model, the scope of applicability and limitations of this reduced model are addressed as follows. 

The main limitations of the model come from the simplified treatment of neutral physics. The influence of neutral particles on the pressure and momentum balance of the SOL plasma is reduced to two 0D parameters, $f_{mom}$ and $f_{pow}$, in the Box model. The complex spatial profile of plasma-neutral power and momentum exchanges and their influence on the SOL plasma is then inherently precluded in the reduced model. Similarly, since the neutral pressure is not included in the pressure balance equation, the Box model is only suited to fully attached SOL conditions. The assumption that all ions are recycled in the SOL as cold neutrals limits the model to situations where the fueling of the main plasma is by recycled particles is not significant, and also overestimates the plasma power lost to the targets. When a close coupling with the 0D neutral module FRANTIC in TRANSP is enabled, fractions of ionization in the core v.s. the SOL and the energies of recycled neutrals can be calculated. Both these last two assumptions will be revised in future iterations of the model so as to form a self-consistent picture with the neutral model.

The Box model also currently only includes a reduced description of the electrons and the main ion species, without an explicit model of impurity particles. This limits the use of the Box model to cases where the effects of the impurity particles are known, for example, from experiments or high fidelity simulations. When impurity effects are significant, such as when impurity puffing or evaporation is used to induce strong radiations, the power loss can be included in the $f_{pow}$ factor.
}

\section{Application to  NSTX Plasma Discharge}\label{sec:NSTX}
\begin{figure}
    \centering
    \includegraphics[width = 0.82\linewidth]{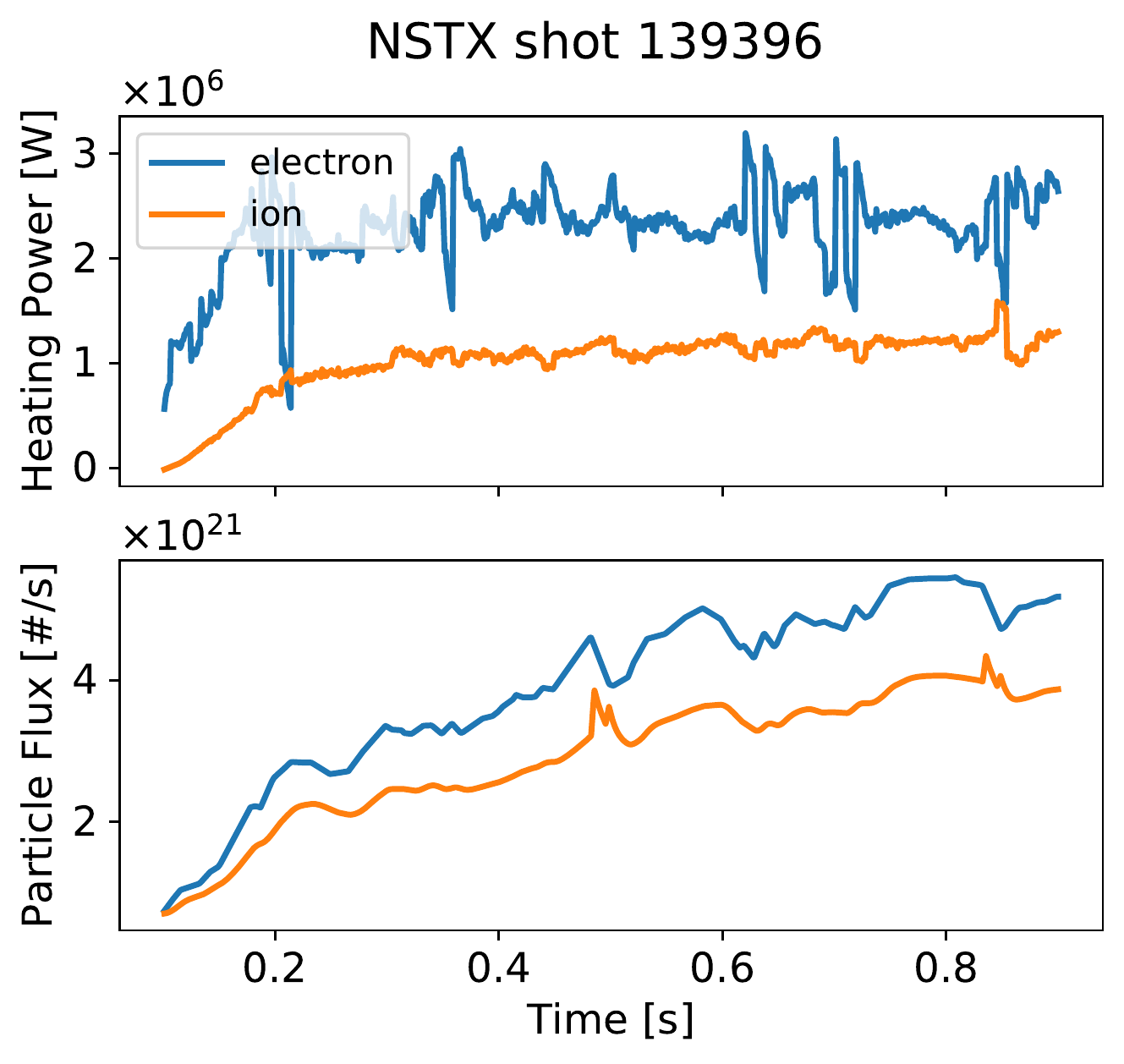}
    \caption{Power and particle fluxes for ions and electrons computed by TRANSP.}
    \label{fig:PF}
\end{figure}

To evaluate the model performance, in this section we compare the plasma parameters calculated from the Box Model against experimental measurements. The shot chosen is NSTX 139396, a lower triangularity, H-mode discharge with solid lithium-coated divertors. Lower recycling \rev{fully attached} divertor conditions are expected because of the lithium coating. The discharge was diagnosed with Multi-Point Thomson Scattering (MPTS) measurements of electron temperature and density \cite{leblanc2003operation}, and Charge-Exchange-Spectroscopy (CHERS) measurements of ion temperature and rotation in the core \cite{bell2010comparison}. At the outer divertor, electron temperature and density are measured with an array of Langmuir probes (LP) \cite{kallman2010high}. Measured profiles are used to constrain the TRANSP simulations, along with EFIT02 equilibria. \rev{The calculated power and particle fluxes, shown in Fig.~\ref{fig:PF}, include effects from the time-dependent evolution of plasma profiles, beam fueling (constrained by data), recycling (calculated from the input particle confinement time and the quasi-neutrality condition), and plasma transport.}

\begin{figure}
    \centering
    \includegraphics[width = 0.82\linewidth]{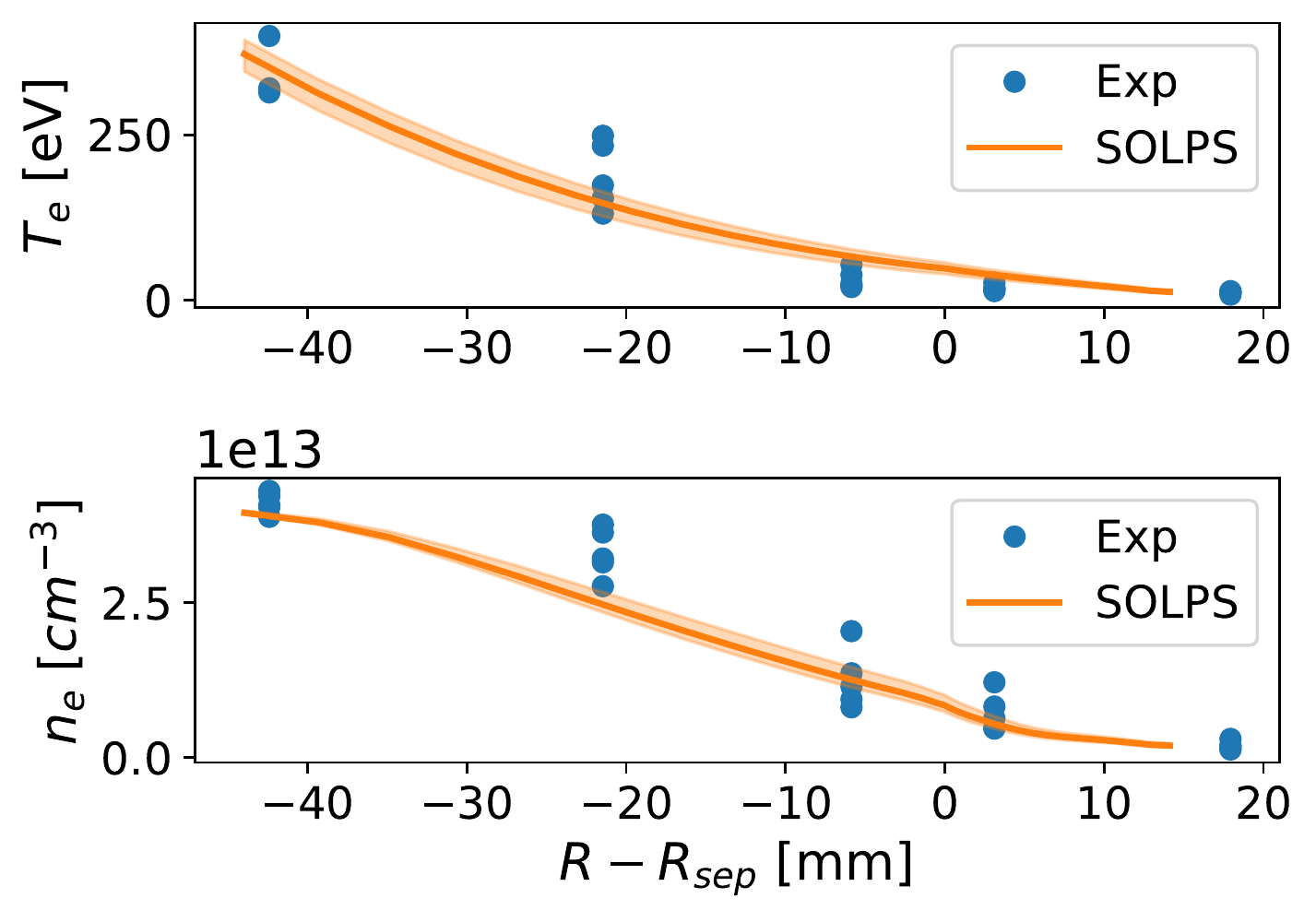}
    \caption{Comparison of plasma electron temperature and density from SOLPS modeling and MPTS measurements. SOLPS modeling is performed with recycling coefficients $0.98 \pm 0.01$, with shaded regions showing the variation of parameters for different recycling coefficients. Experimental measurements for 5 nearby time points are shown.}
    \label{fig:SOLPS_data_comp}
\end{figure}

An interpretive analysis of NSTX shot \#139396 is performed using the SOLPS-ITER code \cite{SOLPS} to constrain the free parameters $f_{pow}$ and $f_{mom}$. The transport coefficients and input parameters for SOLPS are based on past interpretive analyses of this shot, with improved transport coefficients in the SOL to better match MPTS data \cite{emdee2021predictive}. The time slice modeled is $t = 0.331$~s, when the plasma is in ELM-free flat-top without MHD or energetic particle driven instabilties. The heat and particle transport coefficients are iteratively adjusted such that the model reproduces the radial electron temperature and density profiles measured by MPTS. The resulting profiles for recycling coefficients $0.98 \pm 0.01$ are shown in Fig.~\ref{fig:SOLPS_data_comp}. 

\subsection{Calculating Plasma Parameters with Input Recycling}

First we calculate upstream and target conditions with user-specified recycling coefficients. At each step of the TRANSP simulation, the total power and particle flux out of the core is passed to the Box Model. The calculated upstream quantities are compared with the separatrix quantities from TRANSP to serve as a consistency check. The calculated target quantities are compared with Langmuir probe measurements. Connection length is found to be $L = 16$~m via fieldline tracing. In lieu of a neutral model, we set $f_{pow, e} = f_{pow, i} = 0.2$, and $f_R = f_{mom} = 1$, informed by interpretive analysis done with SOLPS-ITER \cite{SOLPS}. The fraction of power conducted is calculated by TRANSP to be approximately $f_{cond} = 0.7$ throughout the discharge.

\begin{figure}
    \centering
    \includegraphics[width = 0.8\linewidth, trim={0 0 0 1cm},clip]{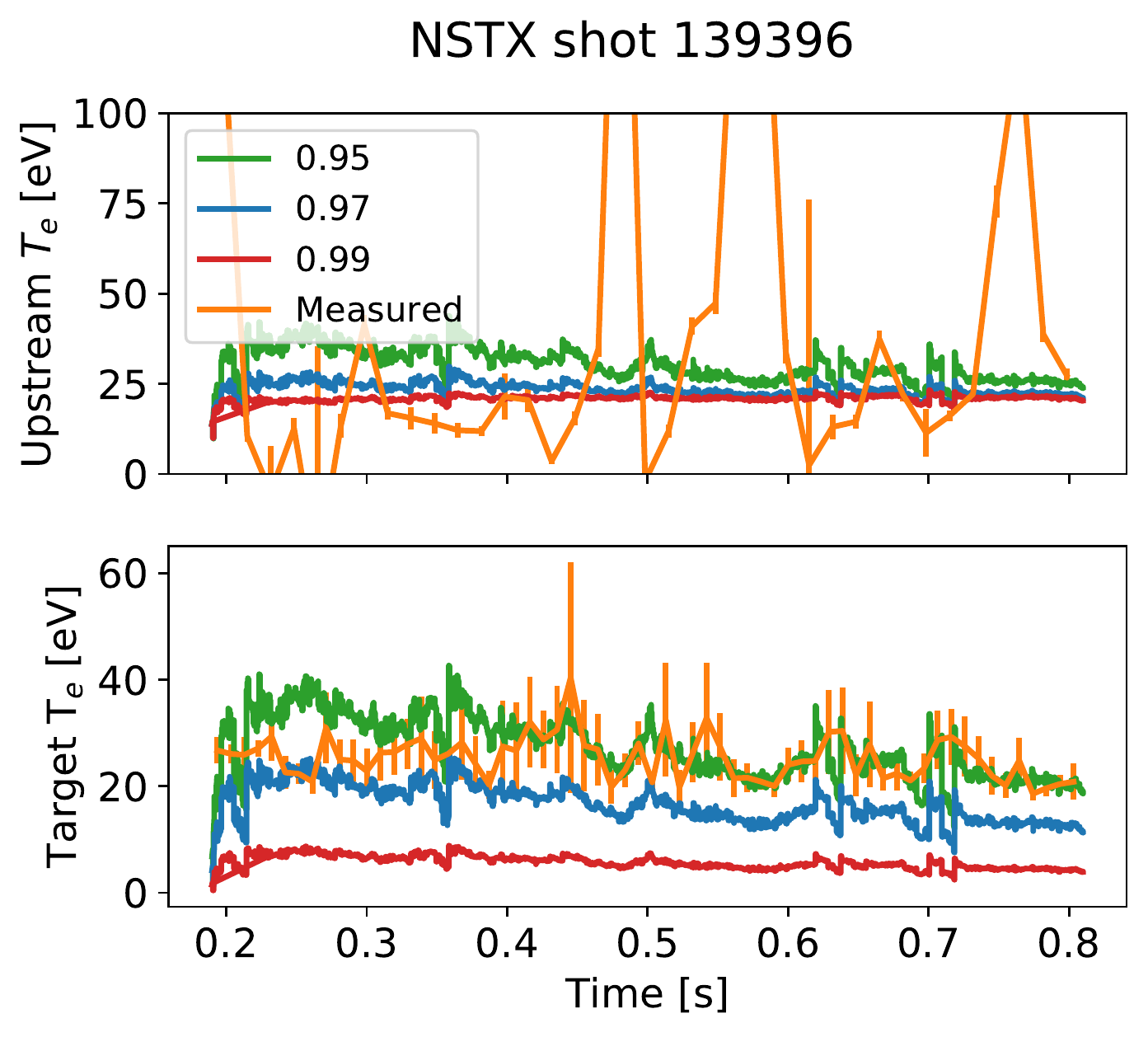}
    \caption{Box model calculated upstream and target electron temperature for 3 different values of recycling coefficients 0.95, 0.97 and 0.99, compared with experimental measurements.}
    \label{fig:Te_rcy}
\end{figure}

\begin{figure}
    \centering
    \includegraphics[width = 0.8\linewidth, trim={0 0 0 1cm},clip]{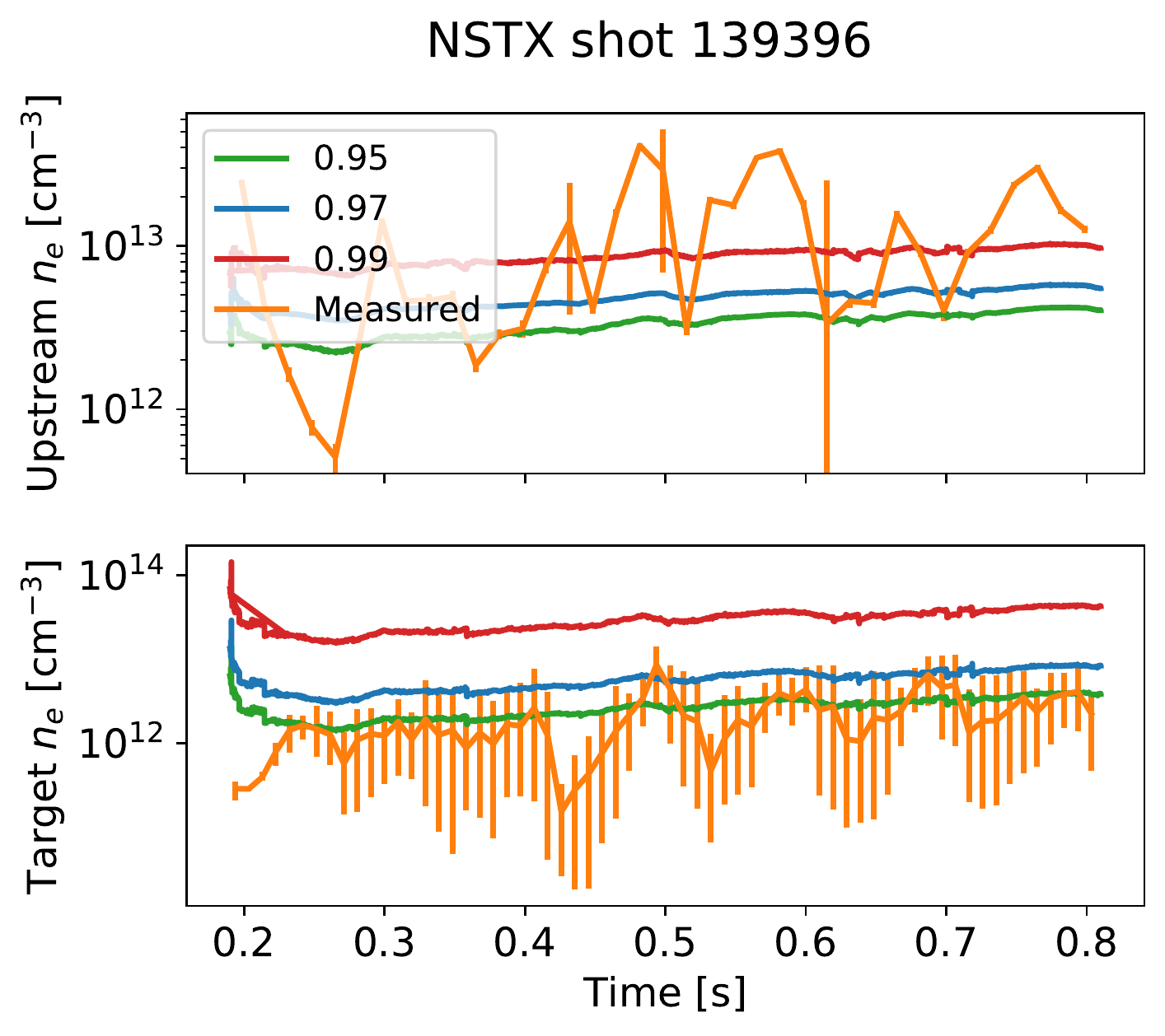}
    \caption{Box model calculated upstream and target electron density for 3 different values of recycling coefficients 0.95, 0.97 and 0.99, compared with experimental measurements.}
    \label{fig:ne_rcy}
\end{figure}

\begin{figure}
    \centering
    \includegraphics[width = 0.8\linewidth, trim={0 0 0 1cm},clip]{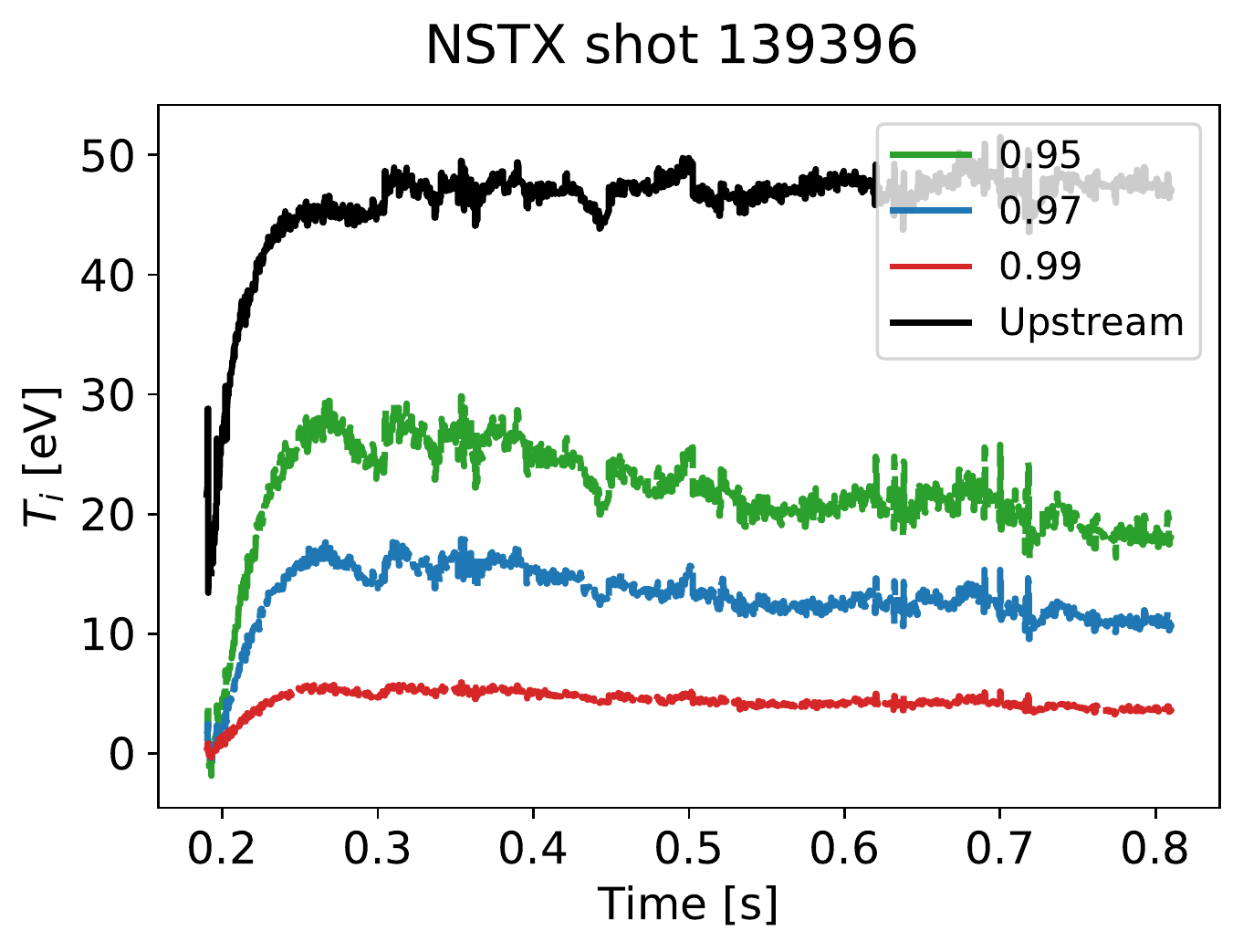}
    \caption{Box model calculated upstream (black) and target (colored) ion temperature for 3 different values of recycling coefficients 0.95, 0.97 and 0.99. Upstream ion temperature is extremely robust and stays the same for the range of recycling coefficients shown here.}
    \label{fig:Ti_rcy}
\end{figure}

Calculated plasma parameters with input recycling coefficients $0.97 \pm 0.02$ are shown in Fig.~\ref{fig:Te_rcy} and Fig.~\ref{fig:ne_rcy}, informed by previous work on similar discharges \cite{maingi2012effect, canik2011edge}. The recycling coefficients for ions and electrons are assumed to be equal. MPTS measurements for electron temperature and density at the separatrix and LP measurements at the target are shown in Fig.~\ref{fig:Te_rcy} and Fig.~\ref{fig:ne_rcy} for comparison. Ion temperatures are also calculated and shown in Fig.~\ref{fig:Ti_rcy}. No experimental measurements for ion temperature are available near the plasma edge in this discharge. Particle confinement times in TRANSP are set at 120~ms. From the electron temperature comparison shown in Fig.~\ref{fig:Te_rcy} and the density comparison shown in Fig.~\ref{fig:ne_rcy}, we can conclude that a recycling coefficient between 0.95 and 0.97 reasonably reproduces observations. Notably, the model-calculated quantities are in general more stable in time than the observed values. 
\rev{Uncertainties in the measured separatrix densities are large because of the limitations of TS measurements in the plasma edge, as evident from the error bar. As can be seen from the edge profiles shown in Fig.~\ref{fig:SOLPS_data_comp}, small uncertainties in the location of the separatrix also corresponds to large differences in the values of temperature and densities.}

\subsection{Inferring Recycling from Target Measurements}
\begin{figure}
    \centering
    \includegraphics[width = 0.8\linewidth, trim={0 0 0 1cm},clip]{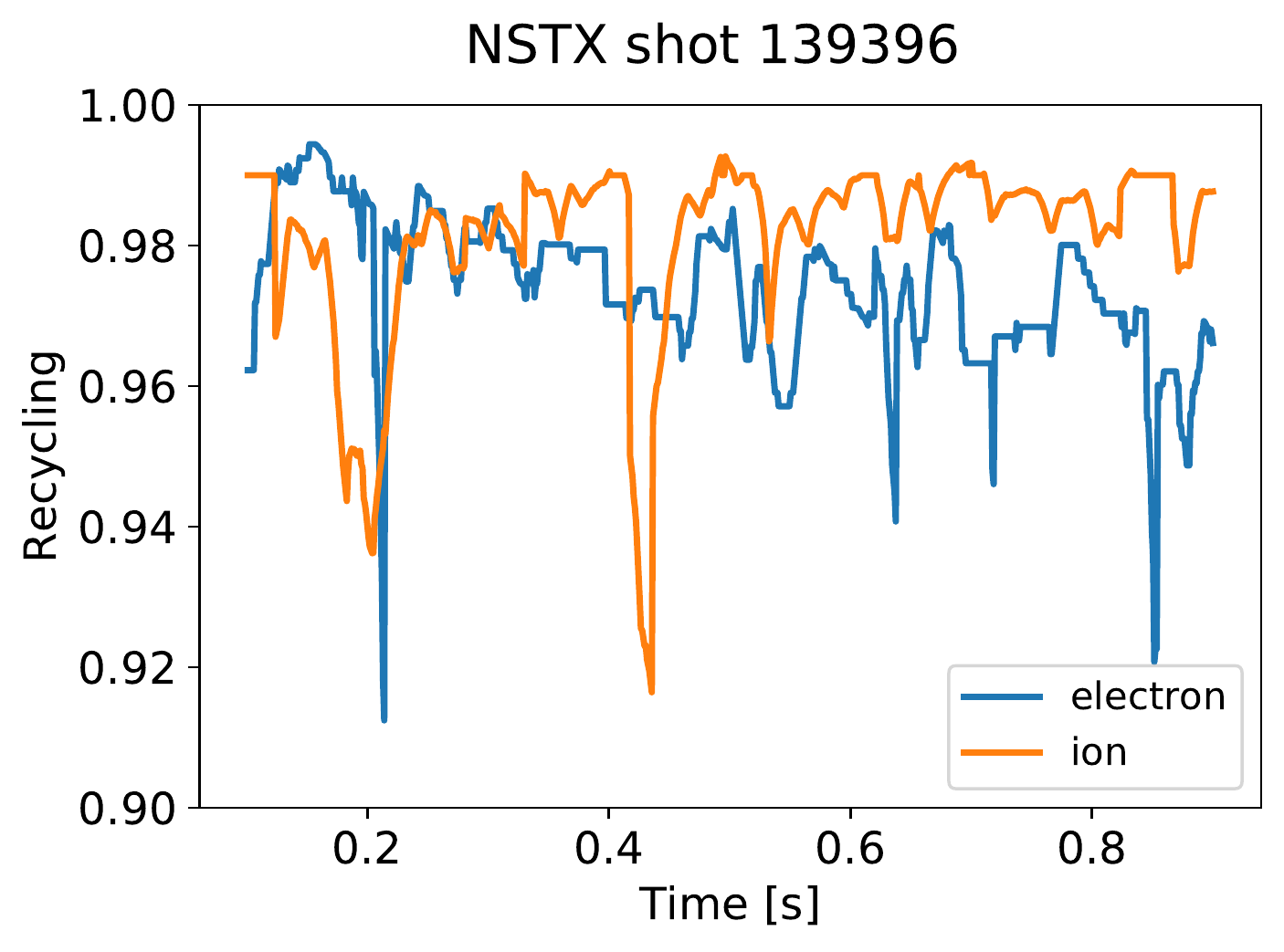}
    \caption{Inferred recycling coefficients for ions and electrons from target LP measurements and TRANSP particle fluxes. The particle confinement time in TRANSP is set at $\tau = 120$~ms.}
    \label{fig:rcy}
\end{figure}
Instead of inputting recycling coefficients, the Box Model can also be constrained by experimental measurements as is commonly done in interpretive analyses. The recycling coefficients can then be calculated and interpreted as outputs of the model. In this section, the Box Model is constrained with LP measurements at the target and the calculated upstream quantities are compared with measurements. Again, the control and free parameters of the model are set at $L = 16$~m, $f_{pow, e} = f_{pow, i} = 0.2$, and $f_R = f_{mom} = 1$. Conduction fraction $f_{cond}$ is calculated by TRANSP time-dependently, and is $\sim 0.7$ throughout the discharge.

\begin{figure}
    \centering
    \includegraphics[width = 0.8\linewidth, trim={0 0 0 1cm},clip]{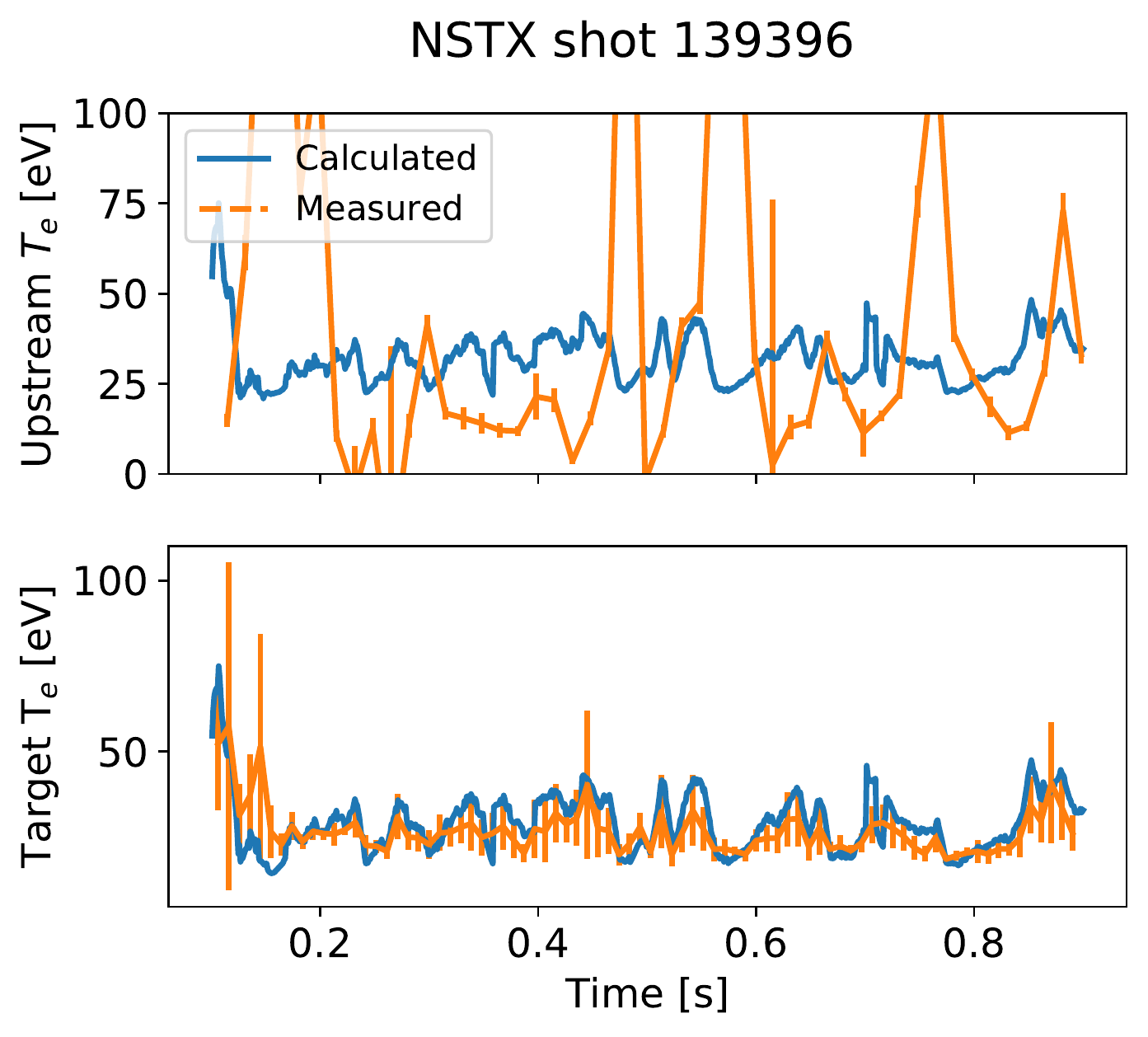}
    \caption{Box model calculated upstream electron temperature compared with measurements at separatrix. The target temperature in the box model is constrained to match LP measurements within error. The calculated upstream temperature is within range of fluctuations of measurements, while being much more stable in time. The large fluctuation in measurements is attributed to uncertainties in the location of the separatrix from EFIT02.}
    \label{fig:Te_int}
\end{figure}

Target electron temperatures measured among the 4 available probes are within error of each other, so the average quantity is used as the representative value. The Box Model target temperature is constrained to be within the range of the 4 measured values. Since the exact location of the strike point is uncertain, the highest target density measured among the 4 probes was used to represent the value of plasma density closest to the strike point. The calculated recycling coefficients from the target constraints are shown in Fig.~\ref{fig:rcy}. Throughout the discharge, the recycling coefficients for both species are found to be consistently below 0.99, which is consistent with expectations from the discharge with solid lithium-coated targets \cite{maingi2012effect, canik2011edge}. Large dips are seen in recycling coefficients, which are attributed to the large dips in measured target electron densities (the most pronounced of which occurs at 0.4~s, see Fig.~\ref{fig:ne_rcy} or Fig.~\ref{fig:ne_int} below). 

Figures \ref{fig:Te_int} and \ref{fig:ne_int} respectively show the calculated electron temperature and density at upstream while target values are constrained to match LP measurements. Similar to the previous section, Fig.~\ref{fig:Te_int} shows that the calculated upstream electron temperature is in general stable in time. The influence on upstream temperature by the target condition is significant in this case since the SOL electrons are almost isothermal. In Fig.~\ref{fig:ne_int}, it is seen that the calculated values of upstream density is within the range of variations of experimental measurements, with a slight overall underestimate. Given the large uncertainties in the target density measured by LP, this underestimate may be tolerated pending further model validations with more precise experimental measurements.

\begin{figure}[h]
    \centering
    \includegraphics[width = 0.8\linewidth, trim={0 0 0 1cm},clip]{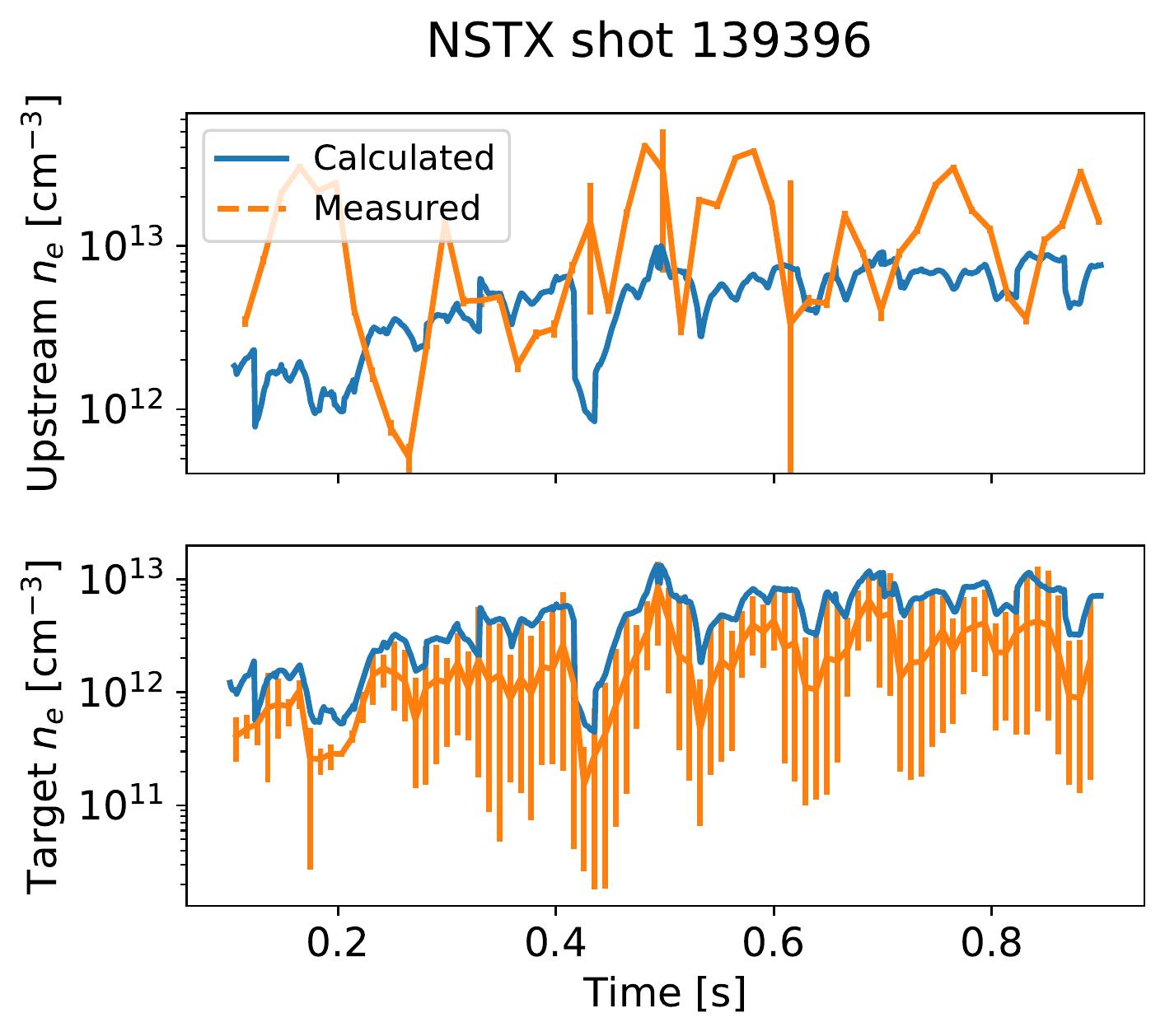}
    \caption{Box model calculated upstream electron density compared with measurements. The target density in the box model is constrained to match the highest observed value among the 4 available LP measurements. The calculated upstream density is within range of fluctuations of measurement, while being more stable in time. }
    \label{fig:ne_int}
\end{figure}

\section{Conclusion}
\label{sec:conclusion}
In this paper we present the SOL Box Model, a reduced model for SOL power and particle balance. In the NSTX plasma example, the Box Model with input recycling coefficients overall produces the measured separatrix electron and ion temperatures within the range of fluctuations, but stable in time. When the simulations are constrained with target LP measurements, the calculated values of recycling coefficients are also consistent with expectations at $<0.99$. 

The SOL Box Model is particularly suited as a reduced model for use in integrated modeling frameworks since it eliminates the need for a fixed upstream density as input. Since the core-to-edge power and particle fluxes are routinely calculated by core transport solvers, the upstream quantities calculated by the Box Model can then be used as boundary conditions in time-dependent iterations. In this scenario, the recycling coefficients are specified by the users, which could be informed by high-fidelity simulations or experimental measurements. The simplicity of the model guarantees little added computational burden to the core solvers, which is especially important for control-room applications, where a fast turn-around time is critical.

\rev{
It is also worth commenting on the role of this reduced SOL model in the context of core-edge coupled modeling. The three boxes, core, SOL, and neutrals, each involve a unique set of physics processes. In a time dependent simulation, the core physics is provided by a transport solver such as TRANSP. What is needed for core-edge integration is the connection from the fluxes out of the separatrix to the vessel wall, which, through SOL and neutral transport processes, feeds back into the core condition. The Box model then serves as the link that connects these core fluxes to the SOL and neutral physics, the latter of which is handled through neutral modeling.

The simple and analytical nature of the Box model first and foremost provides a way to link the core and edge physics without adding extra computational burden. This allows us to test the coupling framework with lower fidelity physics first, before moving on to higher fidelity but also higher cost models. The modularity of the model enables easy replacements of each box when higher fidelity is desired. For example, the neutral model in the 0D box model is reduced to the parameters $f_{pow}$ and $f_{mom}$. These parameters can be found through high fidelity modeling of several time slices, and entered into the Box Model as inputs. To improve the fidelity of the neutrals box, we can couple the box model to a 0D neutral model time dependently (such a model is readily available in TRANSP, and a close coupling of the neutral model with the 0D Box model is currently under testing). Or, to go a step further, we can expand the Box model into two dimensions and couple to a neutral simulation. This capability is also currently under development with DEGAS2 \cite{stotler2000coupling} as the chosen neutral code. The Box Model itself could, in principle, also be replaced by a higher fidelity SOL code. This modularity allows for either fast simulations in the control room, or higher fidelity - and therefore more computationally expensive - simulations for transport studies on shorter time windows within the same framework.
}

\section{Acknowledgement}
This work was supported by US DOE contract number DE-AC02-09CH11466. The authors acknowledge D. Battaglia, R. Maingi, R. Goldston, F. Scotti, and E. Kolmes for helpful discussions, and N. A. Lopez for help preparing the manuscript and general support.

\bibliography{references.bib}

\end{document}